\documentclass[12pt]{article}

\usepackage[utf8]{inputenc}
\usepackage[english]{babel}
\usepackage{helvet}
\usepackage[centertags]{amsmath}
\usepackage{amssymb}
\usepackage{bm}
\usepackage{amsbsy}
\usepackage{graphicx}
\usepackage{appendix}
\usepackage{mathrsfs}
\usepackage{epsfig}
\usepackage{color}
\usepackage{euscript}
\usepackage{ulem}
\usepackage{multirow}
\usepackage{cite}
\newcommand{\fg}{\mathfrak{g}}
\newcommand{\fG}{\mathfrak{G}}

\newcommand{\cH}{\mathcal{H}}
\newcommand{\cK}{\mathcal{K}}
\newcommand{\cL}{\mathcal{L}}

\newcommand{\cP}{\mathcal{P}}
\newcommand{\cU}{\mathcal{U}}

\definecolor{dgreen}{rgb}{0,0.6,0}

\definecolor{darkblue}{rgb}{0., 0, 1}

\definecolor{purple}{rgb}{0.65,0.,0.78}

\definecolor{orange}{rgb}{0.89,0.42,0.05}

\usepackage{jheppubm} 

\newcommand{\be}{\begin{equation}}
\newcommand{\ee}{\end{equation}}
\newcommand{\bea}{\begin{eqnarray}}
\newcommand{\eea}{\end{eqnarray}}

\numberwithin{equation}{section}

\title{Real  Quantum Mechanics in a Kähler Space
}

\author{Igor Volovich}
\affiliation{Steklov Mathematical Institute, Russian Academy of Sciences, \\ Gubkina Street, 8, 119991 Moscow, Russia}

\emailAdd{volovich@mi-ras.ru}

\abstract{In this paper, we demonstrate the equivalence between the complex Hilbert space and real Kähler space formulations of quantum mechanics.

Complex numbers play an important role in the traditional formulation of quantum mechanics in complex Hilbert spaces. However, the necessity of complex numbers—as opposed to their mere convenience—remains a subject of debate. Several alternative formulations of quantum mechanics using real numbers have been proposed. In this paper, we demonstrate that standard quantum mechanics, formulated in a complex Hilbert space, admits an equivalent reformulation in a real Kähler space. By establishing a natural isomorphism between the operator theories of the complex Hilbert space and the real Kähler space, we  prove the equivalence of the two formulations including composite system.

This Kähler-space framework preserves all essential features of quantum mechanics while offering a key advantage: it inherently incorporates a Hamiltonian symplectic structure analogous to classical mechanics. This structural alignment provides a unified geometric perspective for both classical and quantum dynamics. Additionally, we show that the ergodicity of finite-dimensional quantum systems becomes manifest in this framework, resolving interpretational ambiguities present in conventional complex formulations. 
}

\begin{document}

\maketitle
\section{Introduction}

In quantum mechanics, as it was discovered by Heisenberg, Schrödinger, Dirac, and others, complex numbers play a very important role. The mathematical foundations were laid by von Neumann through the theory of operators in a complex Hilbert space \cite{vonNeumann}. 
In particular, the fundamental equation of quantum mechanics, the Schrödinger equation, involves complex numbers. This stands in stark contrast to other foundational equations in theoretical physics, such as Maxwell's and Einstein's equations, which exclusively involve real numbers. The question of why quantum theory requires complex numbers has been a subject of debate for many years \cite{BVN,Varadarajan,Arvind:1995ab,Khrennikov,Khrennikov-book,VVKOGS,Zhu:2020iml,
Hooft,Renou:2021dvp,Chen:2021ril, Li:2021uof,Bednorz:2022zgq,2103.12740,Vedral:2023pij,2407.12755,Takatsuka:2025hez,Sarkar:2025tmd,VEBEQ,Hita:2025okv,Hoffreumon:2025nmq,Feng:2025eci} and refs therein.
It was argued that application of real numbers in physics is an idealization since only rational numbers could be observed. Moreover, the foundamental physicsl laws should be invariant under the change of the number fields \cite{IV-p-adic-numbers}. In particular, one can argue the at the Planck scale non-Archimedean geometry emigres and one has to use p-adic numbers. Quantum mechanics on p-adic numbers has been considerd in \cite{VVZ,DKKVZ}.

The authors of \cite{Renou:2021dvp} argue that while real-number-based quantum theories can replicate outcomes in simple scenarios (e.g., single-system or bipartite experiments), they fail in more complex network configurations involving independent states and measurements, leading to experimentally testable discrepancies. In contrast,  in \cite{Hita:2025okv} it has been 
shown that a physically motivated postulate about composite quantum systems allows to
construct quantum mechanics based on real Hilbert spaces that reproduces predictions for all
multipartite quantum experiments. 

In \cite{Feng:2025eci} it has been claimed that if the independent source assumption holds, then complex-number quantum theory is equivalent to a real-number quantum theory with hidden nonlocal degrees of freedom. This implies that complex numbers are essential for describing entanglement processes between two independent systems. In other words, quantum theory inherently requires complex numbers; otherwise, one would need to adopt a nonlocal real-number quantum theory.
We emphasize that the term real quantum mechanics lacks a universal definition and is interpreted differently across the literature. Many authors define it as a framework rooted in real Hilbert spaces, distinct from the standard complex Hilbert space formalism. Crucially, real and complex Hilbert spaces are not mathematically equivalent, and this foundational distinction gives rise to inequivalent quantum theories.

In this work, we propose a formulation of real quantum mechanics in a real Kähler space, which is equivalent to the standard formulation of quantum mechanics in a complex Hilbert space. Therefore, all results obtained within the framework of the real Kähler space approach will be equivalent to the corresponding results of standard quantum mechanics in the complex Hilbert space approach, including the principle of locality.

In more details, in this paper, we demonstrate that quantum mechanics, usually formulated in the Hilbert space over complex numbers \cite{vonNeumann,Varadarajan,Landau:1991wop}, can be reformulated in the Kähler space over real numbers. 
To a finite-dimensional complex Hilbert space ${\mathbb C}^N$ with inner product $\langle \cdot,\cdot\rangle $ we associate a real Kähler space  consisting of a real Hilbert space ${\mathbb R}^{2N}$ with inner product $g(\cdot,\cdot)$ which is also equipped with a symplectic form $\omega$ and an automorphism $J$. There is a following relation between the  complex Hilbert space and the real Kähler space
\be\label{eq:inner-INT}   \langle \cdot, \cdot \rangle = g(\cdot, \cdot) + i \omega(\cdot, \cdot).
\ee
There is a natural isomorphism in the theories  of operators in the Hilbert space and in the Kähler space, so that 
\be
 \langle  L_1\cdot ...\, L_k \psi,  \, \phi \rangle  = 
   g\Big({\cal L}_1\cdot ...\, {\cal L}_k(q,p),(r,s)\Big) + i \omega\left({\cal L}_1\cdot ...\, {\cal L}_k(q,p),(r,s)\right),\ee
here $L_i$ are operators in the Hilbert space and ${\cal L}_i$ are operators in the Kähler space, see details in Sect.\ref{KHS}. In particular, we associate to each Hermitian, unitary and projection operator in complex Hilbert space their natural partners in real Kähler space. Also the tensor product of complex Hilbert spaces is associated with the tensor product of real Kähler spaces.
 Therefore, quantum mechanics as formulated in the real Kähler space is equivalent to the usual quantum mechanics. 

A Kähler manifold is a manifold where each tangent space is endowed with the structure of a real Kähler space, a complex Hilbert space, or a Hermitian structure. Quantization of Kähler manifolds has been considered in numerous works \cite{Rawnsley:1990tj, Lee:2018czs, Sergeev:2020way}. The difference with the present work is that here we consider quantum theories in real  Kähler spaces.

Quantization of the system with infinite degrees of freedom can be performed directly in the Kähler space. If we have canonical commutation relation in the complex Hilbert space 
\begin{equation}
    [P,Q]=-i,
\end{equation}
we can find an associated operator for 
\begin{equation}
    \mathbb{P}=-J\frac{\partial}{\partial x}, ~~\text{ and }~~~
    \mathbb{Q}=x,
\end{equation}
 and commutation relation
 \begin{equation}
     [ \mathbb{P,Q}]=-J.
 \end{equation}
 The Kähler formulation of quantum mechanics has the advantage over the usual complex formulation of quantum mechanics, since it admits a Hamiltonian symplectic formulation similar to classical mechanics. In particular, the ergodicity of finite-dimensional quantum mechanics becomes transparent.
 
 Specifically, for quantum systems, the Schrödinger equation with an arbitrary quantum Hamiltonian is shown to be equivalent to a classical Hamiltonian system characterized by a quadratic dependence on canonical coordinates and momenta.
 Based on this approach, the ergodicity of finite-dimensional quantum mechanical systems is briefly discussed.

It is well known that the equations of symplectic classical mechanics emerge from the Schrödinger equation in the semi-classical limit \cite{Landau:1991wop}. In this paper, we show that quantum mechanics itself can be represented  as a particular case of symplectic classical mechanics.
Namely, the Schrödinger equation for a quantum mechanical system, 
characterized by an arbitrary quantum Hamiltonian $H$, can be considered as an equation of motion for a classical system with a quadratic Hamiltonian in the form
\be\label{Hlin}
H_{sym} = \frac{1}{2}H_{ab}(p_a p_b + q_a q_b), 
\ee
here $a,b=1,2...$ and summation over repeating indices is assumed.  Here $p_a$ and
$q_a$ are canonical momenta and coordinates, and coefficients $H_{ab}$ are related with the quantum Hamiltonian $H$ of the system.
The classical canonical equations
for the Hamiltonian $H_{sym}$ are equivalent to the Schrödinger equation with the quantum Hamiltonian
$H$.
One obtains the classical Hamiltonian of the same form also for the quantum field theory.

The fact that quantum mechanics can be reformulated as a classical mechanical system with a Hamiltonian that is quadratic in both the canonical momenta and coordinates suggests that, in a formal sense, quantum mechanics effectively reduces to a system of classical harmonic oscillators.
As demonstrated in \cite{VolPadic, Volovich:2019rxl}, any quantum dynamical system is unitarily equivalent to a system of classical harmonic oscillators.

The role of complex and real numbers in quantum mechanics was studied from the early days of quantum theory and still is an active field of research in the so-called real quantum mechanics, see  \cite{BVN, Arvind:1995ab,Khrennikov,VVKOGS,Zhu:2020iml}. The emergence of the quantum mechanics from classical probability theory is considered in \cite{Accardi}. Rewriting the Schrödinger equation for a quantum particle in terms of the real and imaginary parts of the wave function leads to the equivalence of the Schrödinger equation to the Euler-Bernoulli equation in elasticity theory \cite{VEBEQ,Gavrilov:2024mst}.
Reformulation of finite-dimensional quantum mechanics as a symplectic classical mechanics is used to proof the ergodicity of almost all quantum dynamics, see Sect.\ref{sec:Ergod}.   

The paper is organized as follows.
In Sect.\ref{KHS}, we discuss the  theory of real Kähler spaces.  In Sect.\ref{QMK}, we present  formulation of quantum mechanics in real Kähler spaces taking special attention to 
the problem with composite quantum systems. In Sect.\ref{sec:finite}, we derive linear equations governing the real and imaginary parts of the wave function. We then demonstrate that these equations for the real and imaginary components correspond to a Hamiltonian system, with the Hamiltonian explicitly given in Eq.\eqref{Hlin}. In Sect.\ref{sec:Ergod}, we show ergodicity of  finite dimensional quantum systems. 
 In Sect.\ref{sec:QKS}, we
  discuss how to quantize the classical system directly in Kähler space. In Sect.\ref{sec:QFT}, we reformulate quantum field theory using real Kähler space approach.

\section{Kähler and Hilbert Spaces}\label{KHS}  

In this section, we discuss the relationship between Kähler and Hilbert spaces.  
\subsection{ Kähler space}
A Kähler space is a real Hilbert space equipped with a symplectic form and automorphism called complex structure. 
A Kähler space is a quadruplet \((\cK, g,\omega, J)\), where:  
\begin{itemize}  
    \item \(\cK\) is a real vector space,  
      \item \(g: \cK \times \cK\to \mathbb{R}\) is a positive defined the linear form, 
  
    \item \(\omega: \cK \times \cK \to \mathbb{R}\) is a skew-symmetric, non-degenerate bilinear form (i.e., a symplectic form),  
    \item \(J: \cK \to \cK\) is an automorphism satisfying \(J^2 = -\mathrm{id}\) (referred to as a complex structure),
  \item  the following relations hold $g(\cdot, \cdot)=\omega(\, \cdot\,, J\,\cdot\,)$  and $\omega(J\, \cdot\,, J\,\cdot\,)=\omega(\, \cdot\,, \cdot\,),$ 
\end{itemize}  
 see, for example, \cite{Kahler,LKL,Borowiec:2000af}.

One can assume that $(\cK, g)$ is a real Hilbert space and 
$\cK=K\oplus K$, then $J(q,p)=(-q,p)$, $p,q\in K$, here $K$ is a real vector space. In more details this example will be considered below.
\subsection{From Hilbert to Kähler}
If $(\mathcal{H},\langle \cdot, \cdot \rangle)$ is a complex Hilbert space and $\mathcal{H}=H\oplus iH$, where $H$ is a real vector space,  then we can construct the Kähler space.
For $\psi \in \cH$ we set $\psi = q + i p$, where $p,q \in H$ and
\begin{equation}
    \langle \psi_1, \psi_2 \rangle =  \langle q_1 + i p_1,q_2 + i p_2\rangle = \langle q_1,q_2\rangle +\langle p_1,p_2\rangle + i( \langle q_1,p_2 \rangle - \langle p_1,q_2 \rangle)
\end{equation}
Thus we have 
\begin{equation}
    \langle \psi_1, \psi_2 \rangle  = g((q_1,p_1),(q_2,p_2) ) + i \omega((q_1,p_1),(q_2,p_2) ),
\end{equation}
where 
\bea
\label{gomega}g((q_1,p_1),(q_2,p_2) ) &=& \langle q_1q_2\rangle +\langle p_1,p_2\rangle,\\
\omega ((q_1,p_1),(q_2,p_2) ) &=& ( \langle q_1,p_2 \rangle - \langle p_1,q_2 \rangle).
\label{gomega1}
\eea
Therefore, we have the real vector space $\cK=H\oplus H$ and the complex structure $J:\cK \to \cK$ defined by $J(q,p)=(-p,q)$,
and starting from the complex Hilbert space $(\mathcal{H},\langle \cdot, \cdot \rangle)$ we construct the real Kähler space $(\cK, g,\omega, J)$.

\subsection{From Kähler to Hilbert}

Starting from the Kähler space \((\mathcal{K}, g, \omega, J)\), we construct a complex Hilbert space \((\mathcal{H}, \langle \cdot, \cdot \rangle)\).  
Let \(\mathcal{K} = K \oplus K\), and define elements \(\psi \in \mathcal{H}\) as \(\psi = q + i p\), where \(q, p \in K\). The inner product on \(\mathcal{H}\) is given by
\begin{equation}
    \langle \psi_1, \psi_2 \rangle = g\big((q_1, p_1), (q_2, p_2)\big) + i \, \omega\big((q_1, p_1), (q_2, p_2)\big),
\end{equation}
where \(g\) and \(\omega\) are the metric and symplectic form defined in \eqref{gomega} and \eqref{gomega1}, respectively.  

This construction yields a complex Hilbert space, which corresponds to the Hermitian structure on \(\mathcal{K}\).

\subsection{Example}

 Consider the real plane \(\mathbb{R}^2\), which is equivalent to the complex plane \(\mathbb{C}\). We treat \(\mathbb{C}\) as a one-dimensional Hilbert space with the inner product  
\begin{equation}  
    \langle \psi_1, \psi_2 \rangle = (q_1 - i p_1)(q_2 + i p_2) = q_1 q_2 + p_1 p_2 + i\left(q_1 p_2-q_2 p_1   \right),  
\end{equation}  
where \(\psi_a = q_a + i p_a\) for \(a = 1, 2\).  Define the real inner product 
\begin{equation}  
    g((q_1,p_1),(q_2,p_2) ) = q_1 q_2 + p_1 p_2,  
\end{equation}  
and let \(\omega: \mathbb{R}^2 \times \mathbb{R}^2 \to \mathbb{R}\) denote the symplectic form  
\be 
  \omega\left((q_1, p_1), (q_2, p_2)\right) = (q_1, p_1) \Omega \begin{pmatrix} q_2 \\ p_2 \end{pmatrix},  
\ee  
where  
\be  
    \Omega = \begin{pmatrix} 0 & 1 \\ -1 & 0 \end{pmatrix},  
\ee 
and 
\be
\omega\left((q_1, p_1), (q_2, p_2)\right)=
q_1p_2-q_2p_1.
\ee
The mapping \(J: \mathbb{R}^2 \to \mathbb{R}^2\), defined by  
\begin{equation}  
    J = \begin{pmatrix} 0 & -1 \\ 1 & 0 \end{pmatrix},  
\end{equation}  
acts on vectors as \(J(q, p)^T = (-p, q)^T\). This \(J\) is the complex structure on \(\mathbb{R}^2\), completing the Kähler space triplet \((\mathbb{R}^2, \omega, J)\).  

One has 
\be
g(.,.)=\omega( .,J.),
\ee
since $\omega J=I$.

\subsection{Complexification}

Let \( \cK = \mathbb{R}^{2N} = \mathbb{R}^{N} \oplus \mathbb{R}^{N} \) be a real vector space. We define the complexification map \( \gamma: \mathbb{R}^{2N} \to \mathbb{C}^{N} \) by  
\begin{equation}
    \gamma(q, p) = q + i p.
\end{equation}
The complex structure \( J \) on \( V \) is defined via the matrix  
\begin{equation}
    J = \begin{pmatrix} 0 & -I_N \\ I_N & 0 \end{pmatrix},
\end{equation}
where \( I_N \) is the \( N \times N \) identity matrix. The inner product on \( \mathbb{C}^N \) is defined as   
\begin{equation}
    \langle \gamma(\cdot), \gamma(\cdot) \rangle = g(\cdot, \cdot) + i \omega(\cdot, \cdot),
\end{equation}
where \( g \) is the metric and \( \omega \) is the symplectic form. Explicitly, the symplectic form is  
\begin{equation}
    \omega\big((q_1, p_1), (q_2, p_2)\big) = (q_1, p_1) \Omega \begin{pmatrix} q_2 \\ p_2 \end{pmatrix},
\end{equation}
with  
\begin{equation}
    \Omega = \begin{pmatrix} 0 & I_N \\ -I_N & 0 \end{pmatrix}.
\end{equation}
These matrices satisfy \( \Omega J = I_{2N} \), where \( I_{2N} \) is the \( 2N \times 2N \) identity matrix. The metric \( g \) is given by  
\begin{equation}
    g\big((q_1, p_1), (q_2, p_2)\big) = q_1 \cdot q_2 + p_1 \cdot p_2.
\end{equation}

\subsection{Operators}
In this subsection we construct an isomorphism between linear operators on the Hilbert space ${\mathbb C}^N$ and linear operators on the Kähler space $({\mathbb R}^{2N},\omega,J,g)$. 

Let $L: {\mathbb C}^{N} \to {\mathbb C}^{N}$ be a linear operator. We construct an associated operator ${\cal L}: {\mathbb R}^{2N} \to {\mathbb R}^{2N}$ which will be denoted ${\cal L} = \Gamma(L)$  by using the bilinear relation
\be\label{Gamma}
 \langle \gamma(\cdot), L\gamma(\cdot) \rangle = g(\cdot, \cL\cdot) + i \omega(\cdot, \cL\cdot).
\ee
If there is an operation, for example,  $*$ acting on the operator $L$ in the Hilbert space
\be
 \langle \gamma(\cdot), L^*\gamma(\cdot) \rangle =  \langle L\gamma(\cdot), \gamma(\cdot) \rangle .
\ee
then the corresponding operation acting on the operator $\cL$ on the  Kähler space, i.e. $\cL^*$ is defined as follows
\bea
g(\cdot, \cL^*\cdot) &=&g(\cL\cdot, \cdot),\\
\omega(\cdot, \cL^*\cdot) &=&\omega(\cL\cdot, \cdot).\eea

 We represent $L$ as $L = X + iY$, where $X$ and $Y$ are real matrices. We write
\[
L\psi = (X + iY)(q + ip) = Xq - Yp + i(Yq + Xp).
\]
Therefore, we define ${\cal L}$ as a block matrix
\[
{\cal L} = \begin{pmatrix} X & -Y \\ Y & X \end{pmatrix}.
\]
We have relations between expectation values of operators on the Hilbert and Kähler spaces:
\[
\langle \psi, L\phi \rangle = g\left({\cal L}(q,p), (r,s)\right) + i \omega\left({\cal L}(q,p), (r,s)\right),
\]
where 
\[
\psi = q + ip = \gamma(q,p), \qquad \phi = r + is = \gamma(r,s).
\]
We also have the relation 
$$
\langle L_1 \cdots L_k \psi, \phi \rangle = g\left({\cal L}_1 \cdots {\cal L}_k(q,p), (r,s)\right) + i \omega\left({\cal L}_1 \cdots {\cal L}_k(q,p), (r,s)\right).
$$

\subsection{Kähler unitary group $U(N)$ and symplectic group $Sp(2N, {\mathbb R})$}
In this section, for a unitary operator in the Hilbert space $\mathbb{C}^N$, we define an associated operator $\mathcal{U} = \Gamma(U)$, which will be called a Kacher unitary operator.  
In fact, we show that the unitary group $U(N)$ is a subgroup of the symplectic group $Sp(2N, \mathbb{R})$. Let us recall that a symplectic matrix $S$ of the symplectic group $Sp(2N, \mathbb{R})$ satisfies the relation
\begin{equation}
    S^T J S = J,
\end{equation}
where 
 \be J = \left( \begin{array}{ccc} 0~~~I\\-I~~~0 \end{array}\right).\ee 

Let $S$ be a $2N \times 2N$ matrix in the form
\begin{equation*}
S = \left(\begin{array}{ll}
A & B \\
C & D
\end{array}\right),
\end{equation*}
where $A, B, C, D$ are $N \times N$ matrices. 
The conditions on $A, B, C$, and $D$ are:

\begin{itemize}
    \item $A^{\mathrm{T}} C$, $B^{\mathrm{T}} D$ symmetric, and $A^{\mathrm{T}} D - C^{\mathrm{T}} B = I$;
    \item $A B^{\mathrm{T}}$, $C D^{\mathrm{T}}$ symmetric, and $A D^{\mathrm{T}} - B C^{\mathrm{T}} = I$.
\end{itemize}

Let us show that there is a canonical mapping $\gamma$ of the unitary group $U(N)$ to $\mathrm{Sp}(2N, \mathbb{R})$. Let $V \in U(N)$. We present $V$ in the form 
\begin{equation}
    V = X + iY, \quad \text{where } X, Y \text{ are } N \times N \text{ matrices with real entries}.
\end{equation}
We define $\gamma$ by the following formula:
\begin{equation} \label{gamma}
    \gamma(V) = \gamma(X + iY) = 
    \left(\begin{array}{ll}
X & Y \\
-Y & X
\end{array}\right).
\end{equation}
One can check that the unitarity conditions $VV^* = V^*V = I$ lead to the conditions for the symplectic matrices.

One has an important relation:
\begin{equation}
    U(N) = \mathrm{Sp}(2N, \mathbb{R}) \cap \mathrm{O}(2N, \mathbb{R}),
\end{equation}
where $U(N)$ is the unitary group, $\mathrm{Sp}(2N, \mathbb{R})$ is the symplectic group, and $\mathrm{O}(2N, \mathbb{R})$ is the orthogonal group. We recall that the symplectic group is defined as the group of linear transformations $S$ on $V$ that satisfy the relation $\omega(S\chi, S\eta) = \omega(\chi, \eta)$ for all $\chi, \eta \in V$.

\subsection{Hermitian operators}
Here for a Hermitian operator $L$ in the Hilbert space $\mathbb{C}^N$, we associate a $K$-Hermitian operator $\mathcal{L}$ on the Kähler space $(\mathbb{R}^{2N}, \omega, J, g)$. The condition of Hermiticity
\begin{equation}
    \langle \psi, L\phi \rangle = \langle L\psi, \phi \rangle
\end{equation}
now reads
\begin{align}
    g\left(\mathcal{L}(q,p), (r,s)\right) &= g\left((q,p), \mathcal{L}(r,s)\right) \label{eq:g_condition} \\
    \omega\left(\mathcal{L}(q,p), (r,s)\right) &= \omega\left((q,p), \mathcal{L}(r,s)\right) \label{eq:omega_condition}
\end{align}
Thus we have
\begin{equation}
    \mathcal{L}^{\mathrm{T}} = \mathcal{L}, \quad \mathcal{L}\Omega = \Omega\mathcal{L}. \label{eq:properties}
\end{equation}

Let $(\mathcal{H}, \langle \cdot, \cdot \rangle)$ be a complex Hilbert space and the quadruplet $(V, \omega, J, g)$ be the associated Kähler space obtained by setting $\psi = q + i p$.

Let represent the  Hermitian operator $L$ as a pair of real operators:
\begin{equation}
    L = A + iB \label{eq:decomposition}
\end{equation}
where $A$ and $B$ are real symmetric and skew-symmetric operators respectively. We define the mapping $\gamma$ from Hermitian operators on $\mathcal{H}$ to operators on the Kähler space $V$:
\be
   \gamma(L) = \begin{pmatrix} A & B \\ -B & A \end{pmatrix} \label{eq:gamma_matrix}
   \ee 
   and $\gamma(L)$ satisfy the relations 
   \bea
g\left(\gamma(\mathcal{A})\begin{pmatrix} p \\ q \end{pmatrix}, \begin{pmatrix} r \\ s \end{pmatrix}\right) &=& g\left(\begin{pmatrix} p \\ q \end{pmatrix}, \gamma(\mathcal{A})\begin{pmatrix} r \\ s \end{pmatrix}\right) \label{eq:g_preserve} \\
\omega\left(\gamma(\mathcal{A})\begin{pmatrix} p \\ q \end{pmatrix}, \begin{pmatrix} r \\ s \end{pmatrix}\right) &=& \omega\left(\begin{pmatrix} p \\ q \end{pmatrix}, \gamma(\mathcal{A})\begin{pmatrix} r \\ s \end{pmatrix}\right) \label{eq:omega_preserve}
\eea
  where
  $\psi = q + ip, \quad \phi = r + is$.

\subsection{Tensor Products}\label{TensorProd}
Here we consider two natural concepts of tensor products for real Kähler spaces.
\begin{itemize}
    \item 1. {\it Tensor product over real number fields:} $\mathbb{R}^{2m}\otimes_{\mathbb{R}}\mathbb{R}^{2n}
    =\mathbb{R}^{4mn}$.
      \end{itemize}  
    One can define  the tensor product of two  Kähler space just using that  Kähler space is a real space and consider the tensor product of two  Kähler space as a product of two spaces over real number fields. 
In this case we deal with
     \be \mathbb{R}^{2m}\otimes_{\mathbb{R}}\mathbb{R}^{2n}
    =\mathbb{R}^{4mn}\ee
 \begin{itemize}   
 \item 2. {\it Tensor product over complex number fields:} $\mathbb{R}^{2m}\otimes_{\mathbb{C}}\mathbb{R}^{2n}
    =\mathbb{R}^{2mn}$.
    \end{itemize}
    One can define the tensor product of two Kähler spaces using their relation with  Hilbert spaces.
    In this case,
to construct the tensor product of two Kähler spaces \be\xi_A = (\mathbb{R}^{2m}_A, g_A, \omega_A, J_A)\ee and 
\be\xi_B = (\mathbb{R}^{2n}_B, g_B, \omega_B, J_B),\ee we first transform them into complex Hilbert spaces $(\mathbb{C}_A^m, \langle \cdot, \cdot \rangle_A)$ and $(\mathbb{C}_B^n, \langle \cdot, \cdot \rangle_B)$. We then form the tensor product of these complex Hilbert spaces:
\begin{equation}
(\mathbb{C}_A^m,\langle \cdot, \cdot \rangle_{A}) \otimes (\mathbb{C}_B^n,\langle \cdot, \cdot \rangle_{B}) = (\mathbb{C}_{AB}^{mn}, \langle \cdot, \cdot \rangle_{AB}),
\end{equation}
and transform the resulting complex Hilbert space back into a real Kähler space:
\begin{equation}
\label{xiAB}
\xi_{AB} = (\mathbb{R}^{2mn}_{AB}, g_{AB}, \omega_{AB}, J_{AB}).
\end{equation}
This yields the real Kähler space $\xi_{AB}$ describing the composite system consisting of $\xi_A$ and $\xi_B$. We denote this operation as the tensor product of real Kähler spaces over the complex number field:
\begin{equation}\label{CC}
\xi_{AB} = \xi_A \otimes_{\mathbb{C}} \xi_B.
\end{equation}

Note that here an application of complex numbers is only for convenience, to make the exposition shorter, and in fact one can deal with the real numbers only.

\section{Quantum Mechanics in Real Kähler Space}
\label{QMK}

In this Section, we demonstrate that quantum mechanics, usually formulated in the Hilbert space over complex numbers \cite{vonNeumann,Varadarajan,Landau:1991wop}, can be reformulated in the Kähler space over real numbers. 
To a finite-dimensional complex Hilbert space ${\mathbb C}^N$ with inner product $\langle \cdot,\cdot\rangle $ we associate a real Kähler space  consisting of a real Hilbert space ${\mathbb R}^{2N}$ with inner product $g(\cdot,\cdot)$ which is also equipped with a symplectic form $\omega$ and an automorphism $J$. There is a following relation between the  complex Hilbert space and the real Kähler space
\be\label{eq:inner-INT}   \langle \cdot, \cdot \rangle = g(\cdot, \cdot) + i \omega(\cdot, \cdot).
\ee
There is a natural isomorphism in the theories  of operators in the Hilbert space and in the Kähler space, so that 
\be
 \langle  L_1\cdot \cdot\cdot\, L_k \psi,  \, \phi \rangle  = 
   g\Big({\cal L}_1\cdot \cdot\cdot\, {\cal L}_k(q,p),(r,s)\Big) + i \omega\left({\cal L}_1\cdot ...\, {\cal L}_k(q,p),(r,s)\right),\ee
here $L_i$ are operators in the Hilbert space and ${\cal L}_i$ are operators in the Kähler space, see the previous Sect. \ref{KHS}. In particular, we associate to each Hermitian, unitary and projection operator in complex Hilbert space their natural partners in real Kähler space. Also the tensor product of complex Hilbert spaces is associated with the tensor product of real Kähler spaces.
 Therefore, quantum mechanics as formulated in the real Kähler space is equivalent to the usual quantum mechanics. 
 
 We can formulate the main principles of quantum mechanics directly in the real Kähler space, compare with  corresponding formulation \cite{SevenPrinc,Ohya} in the Hilbert space.

 \subsection{Real Kähler space}

To a physical system one assigns a real Kähler space $\cK$ (instead of the a Hilbert space $\cH$). The
observables correspond  to the  real Kähler self-adjoint operators in ${\cK}.$ The pure states correspond  to the one-dimensional subspaces
of ${\cK}.$ In the finite dimensional Kähler space $\cK={\mathbb R}^{2n}$ the expection value of the 
Kähler Hermitian operator
\be
\cL=\begin{pmatrix} A & B \\ -B & A \end{pmatrix},\ee
is
\bea
g\left(\begin{pmatrix} p & q\end{pmatrix}, \cL\begin{pmatrix} p \\ q\end{pmatrix}\right)
=g(q,Aq)+g(p,Ap)+2g(p,Bq)
   \eea

\subsection{Measurements}

Measurement is an external intervention which changes the state of
the system. Since the measurements in usual Hilbert space formulation of quantum mechanics  are dicsribed by the projector operator, in the real Kacher formulation they are described by operators $\cP$ associated with projectors in the Hilbert space,
\be
\cP=\Gamma(P),\ee
where $P$ is a projector and $\Gamma$ as in \eqref{Gamma}.

\subsection{Time}

The dynamics of the vector $(q,p)$ in the Kähler space $\cK$ is given by one parameter symplectic orthogonal group and of a state $\psi$
in the  Kähler space which occurs with passage of time is given by

\be
 \begin{pmatrix} p (t)\\ q(t)\end{pmatrix}=\cU(t)\begin{pmatrix} p \\ q\end{pmatrix}
\ee
The infinitesimal form of this dynamics is the Hamiltonian equations 
\be
\dot q(t)=\frac{\partial H_{sym}}{\partial p},
\quad \dot p(t)=-\frac{\partial H_{sym}}{\partial q}\ee
where $H_{sym}$ is quadratic inmomenta and coordinates. 

\subsection{Space}
All physical processes occur in three-dimensional Euclidean space
${\mathbb R}^3$.
Its group of motion is formed by the translation group $T^3$ and
the rotation group $O(3)$. In the  Kähler space
there is a symplectic orthogonal representation $\cU(a)$ of the
translation by the three-vector $a.$  The generators of the translation group are $\cP_j=-J\partial/\partial x_j $, $j=1,2,3$. Also there are representations of the the rotation group $O(3)$.

\subsection{Internal symmetries}

There is a compact Lie group $\fG$ of internal symmetries and
its symplectic orthogonal  representation $\cU(\fg), \fg \in \fG$ in the
Kähler space ${\cK}$ which commutes with representations of
the translation group $\cU(a)$ and the rotation group. For
instance one could have the gauge group $\fG=U(1)$ which
describes the electric charge.
The group generates the superselection sectors. In the case of the Minkowski space one can introduce the principle of locality in quantum field theory \cite{SevenPrinc,Ohya}.\\

\subsection{Composite Systems}

If there are two different systems with assigned the real Kähler spaces
$\cK_A$ and $\cK_B$ then the composite system is
described by the tensor product of them. As has been discussed in Sect. \ref{TensorProd}, there are two different definitions of the tensor products. Here we mean 
 the tensor product of real Kähler spaces over the complex number field \eqref{CC},
\be\label{CCm}
\cK_{AB}=\cK_A\otimes_{\mathbb C}\cK_B.
\ee

Let us emphasize that the symbol $\mathbb C$ 
 on the right-hand side of \eqref{CCm} is introduced purely for notational convenience; all quantities in this context are strictly real-valued.

Let us note that the problem with composite quantum systems  based on real Hilbert space  is widely  discussed in the literature \cite{Renou:2021dvp,Hita:2025okv,Hoffreumon:2025nmq,Feng:2025eci}. We show how this problems can be solved in the approach with the real Kähler space formulation. 

In \cite{Renou:2021dvp}, it is shown that real and complex quantum theory make different predictions in network scenarios comprising independent states and measurements. This allows one to devise a Bell-like experiment whose successful realization would disprove real quantum theory.
However, in \cite{Hita:2025okv} it has been shown that a physically motivated postulate about composite quantum systems permits the construction of quantum mechanics based on real numbers that reproduces predictions for all multipartite quantum experiments. 

We note that in this debate, careful attention must be paid to what is meant by ``quantum mechanics based on real numbers.'' Often, real quantum mechanics is understood merely as a mathematical formulation in a real Hilbert space. In the present paper, we use the Kähler space formulation of real quantum mechanics. A Kähler space includes not only a real Hilbert space but also a symplectic form $\omega$ and an automorphism $J$, as discussed in the previous section. The crucial property of a real Kähler space is its isomorphism to a complex Hilbert space. Therefore, quantum mechanics constructed via the real Kähler space formulation is automatically equivalent to standard quantum mechanics formulated in a complex Hilbert space, including the standard tensor product for composite systems.

\subsection{Bose-Fermi alternative}

The Kähler space of an $N$-particle system is the $N$-fold tensor
product of the single particle Kähler spaces provided that the
particles are not of the same species. For identical particles
with integer spin (bosons) one uses the symmetrized $N$-fold
tensor product $({\cK}^{\otimes N})_S$  of the Kähler space
${\cK}.$ For identical particles with half integer spin
(fermions) one uses the anti-symmetrized $N$-fold tensor product
$({\cK}^{\otimes N})_A$.

\section{Dynamics in Kähler Space} \label{sec:finite}
\subsection{Schrödinger equation in a finite-dimensional Kähler space}
Consider a quantum system described by a finite-dimensional Kähler space \(\mathcal{H} = \mathbb{C}^N\). The evolution is governed by the Schrödinger equation
\begin{equation} \label{Schreq}
  i\, \dot{\psi}_a = H_{ab}\, \psi_b, \quad a, b = 1,\dots,N,
\end{equation}
where \(\psi_a \in \mathbb{C}\) and \(H_{ab}\) is a Hermitian matrix (i.e., \(H_{ab} = \overline{H}_{ba}\)). By expressing the complex wave function in terms of its real and imaginary components 
\begin{equation} \label{psi_decomp}
  \psi_a = q_a + i\, p_a,
\end{equation}
and writing the Hamiltonian as
\begin{equation} \label{H_decomp}
  H_{ab} = K_{ab} + i\, L_{ab},
\end{equation}
with \(K_{ab} = K_{ba}\) (symmetric) and \(L_{ab} = -L_{ab}\) (antisymmetric), equation~\eqref{Schreq} decomposes into two  real equations:
\begin{equation} \label{qdyn}
  \dot{q}_a = K_{ab}\, p_b + L_{ab}\, q_b,
\end{equation}
\begin{equation} \label{pdyn}
  \dot{p}_a = -K_{ab}\, q_b + L_{ab}\, p_b.
\end{equation}

Equations \eqref{qdyn} and \eqref{pdyn} are immediately recognized as the canonical Hamiltonian equations
\begin{equation}
  \dot{q}_a = \frac{\partial H_{sym}}{\partial p_a}, \quad \dot{p}_a = -\frac{\partial H_{sym}}{\partial q_a},
\end{equation}
with the quadratic Hamiltonian function
\begin{equation} \label{H_classical_finite}
  H_{\text{cl}} = \frac{1}{2} \left(p_a\, K_{ab}\, p_b + q_a\, K_{ab}\, q_b \right) + p_a\, L_{ab}\, q_b.
\end{equation}
Thus, the evolution of the finite-dimensional quantum system is completely equivalent to that of a classical system with the specified quadratic Hamiltonian.

\subsection{Ergodicity}\label{sec:Ergod}
In this section, we briefly mention how the approach to quantum mechanics as classical mechanics can be used to show ergodicity of almost all finite dimensional quantum systems. Definition.
A finite-dimensional quantum dynamical system is called ergodic if the associated classical dynamical system 
$\{ M, \varphi_t, \mu\}$ on the surface of the integrals of motion is ergodic.

Let us remind that ergodicity for classical system means  
the time average for an integrable function $f$ coincides with the spatial average almost everywhere
\begin{equation}
\lim _{T \to \infty}\frac{1}{T}\,
\int _0^T f(\varphi_t(x)) dt = \int _{M} f d\mu.
\end{equation}

The constants of motion  \begin{equation}
    F_a = \xi_a^2 + \eta_a^2, \quad a=1,...N. \end{equation}

 In the action-angle variables  a linear flow on a torus:
$\dot J_a = 0, \quad \dot \theta_a = \lambda_a \ (\text{mod } 1)$,\\ $a=1,...N$.

{\bf Theorem}. A finite-dimensional quantum dynamical system with rationally independent eigenvalues of the Hamiltonian is ergodic.

 Thus, in a finite-dimensional Hilbert space, almost any quantum dynamical system is ergodic.

 \section{Quantization in the Kähler space}\label{sec:QKS}
 Let us discuss how to quantize the classical system directly in the Kähler space.

To the this end we consider the simplest case of the quantum system described in the  Hilbert space $L^2(\mathbb{R})$. 
We have the operator of position $Q$ as the multiplication on  $x$, 
\begin{equation}
    (Q\psi)(x)=x(q(x)+ip(x))
\end{equation} 
and the momentum operator 
\be
(P\psi)(x)=-i\frac{d}{d x}\left(q(x)+ip(x)\right)=\frac{d}{d x}(p(x)-iq(x)).
\ee
Therefore for the associated operators in the Kähler space we have,
\bea
    \mathbb{Q}(q(x),p(x))&=&x(q(x),p(x))=(xq(x),xp(x))
\\
    \mathbb{P}(q(x),p(x))&=& - J\frac{d}{d x}(q(x),p(x))= - J(\frac{d}{d x}q(x),\frac{d}{d x}p(x))
\eea
Therefore we have the communation relation in the Kähler space
 \begin{equation}
     [ \mathbb{P,Q}]=-J.
 \end{equation}

One can define the $C^*$
 and von Neumann algebras on the real Kähler space by using the corresponding structure of the real Hilbert space.

Consider the Schrödinger equation 
\bea
\label{3.1}
    i \hbar \dot{\psi}& =& H \psi,
    \\
    H & =& - \frac{\hbar^2}{2m} \Delta+V(x),\label{Hdelta}
\label{3.2}\eea
where $\Delta$ is the Laplace operator in $\mathbb{R}^n$ and $V(x)$ is a potential, $\hbar$ the Planck constant.

Decomposing the wave function into its real and imaginary parts as
\begin{equation}
  \psi(t,x) = q(t,x) + i\, p(t,x),
\end{equation}
we obtain from  \eqref{3.1} the following equations 
\begin{equation}
  \dot{q}= \hbar^{-1} H p,
\qquad
\dot{p} = -\hbar^{-1} Hq,
\end{equation}
which are canonical equations for the classical Hamiltonian  $H_{sym}$,
\begin{equation}\label{Hsym}
    H_{sym}=\frac{1}{2\hbar} \int( pHp+qHq)dx.
\end{equation}

Note that the form of the Hamiltonian
\eqref{Hsym} with H given by \eqref{Hdelta} follows the quantization of the classical particle in the Kähler space. Indeed,
the classical particle Hamiltonian has the form
\be
H_{part}=\sum _{k=1}^3 \frac{p_k^2}{2m}+V(x)\ee
Quantization in the Kähler space consists in substitution 
\be
p_k\to \cP_k=-J\hbar\frac{\partial }{\partial x_k}\ee
and we get
\be
H_{part}\to  -\frac{\hbar ^2}{2m}\Delta+V(x)\ee

Let $\{e_a\}$, where $a=1,2,...$, be a basis in 
$L^2(\mathbb{R}^n)$.  We can then express $q$ and 
$p$ as: 
\begin{equation}
    q=\sum_{a=1}^\infty q_a(t) e_a(x),
\qquad
    p=\sum_{a=1}^\infty p_a(t) e_a(x)
\end{equation}
and the classical Hamiltonian $H_{sym}$ takes the following form:
\begin{equation}
    H_{sym}=\frac{1}{2\hbar}\sum_{a,b=1}^{\infty} H_{ab}(p_ap_b+q_aq_b) ,
\end{equation}
where
\begin{equation}
H_{ab} = \int e_a H e_b\, dx.
\end{equation}
This Hamiltonian represents the limit of a finite-dimensional Hamiltonian:

\begin{equation}
    H_{cl } = \lim_{N\to\infty} H_{cl,N}\,,
\end{equation}
where 
\begin{equation}
    H_{cl,N}=\frac{1}{2\hbar}\sum_{a,b=1}^{N} H_{ab}(p_ap_b+q_aq_b). 
\end{equation}
Therefore, the Schrödinger equation in infinite-dimensional Hilbert space $L^2(\mathbb{R}^n)$
is reformulated as a classical mechanical system.

One can use a linear canonical transformation to diagonalize $H_{sym}$ and bring it to the following  diagonal form: 
\begin{equation}
    H_{sym} =\frac{1}{2} \sum_a \lambda_a (p_a^2+q_a^2).
\end{equation}
To obtain this form, one can solve the eigenvalue problem for the quantum Hamiltonian:
\begin{equation}
        H e_a=\lambda_ae_a,
    \end{equation}
 which shows that the problem of diagonalizing the classical Hamiltonian is equivalent to solving the eigenvalue problem for the quantum Hamiltonian.

Additionally, the following formula holds:  
    \begin{equation}
        H_{sym}=\frac{1}{2} \langle \psi | H | \psi \rangle . 
    \end{equation}

    {\bf Remark 1}. We refer to 
$H_{sym}$
as the classical Hamiltonian. However, it is important to emphasize that 
$H_{sym}$
  depends on Planck's constant and should not be confused with the classical Hamiltonian, which is given by
 \be
    H_{\mbox{classical}}
    =\frac{1}{2m}p^2 +V(x)\ee
\section{Quantum Field Theory in Kähler Space} \label{sec:QFT}
Consider scalar quantum field theory with the Hamiltonian 
\begin{equation}
    H= \int \left(\frac{1}{2} \pi^2(x) + \frac{1}{2} \nabla \phi(x) + V(\phi(x)) \right) d^nx,
\end{equation}
where
\begin{equation}
    [\pi(x),\phi(y)]=-i\delta(x-y).
\end{equation}

We assume that the wave function is the functional on $\phi$, $\psi=\psi[\phi]$ and \\$\pi(x)=-i \delta/\delta \phi(x)$. We consider path integrals with respect to the formal measure ${\cal D}\phi$ and the space with $\int  | \psi| ^2 {\cal D}\phi<\infty $. The volume and ultraviolet regularizations are assumed.  Let $e_a[\phi]_{a=1}^\infty $ be a basis in this space. Thus, the treatment of quantum field theory as a classical mechanical system is reduced to the framework discussed in the previous section.

This implies that quantum field theory can be reformulated as a classical mechanical system with a Hamiltonian 
 \begin{equation}
    H_{sym}=\frac{1}{2}\sum_{a,b=1}^{\infty} H_{ab}(p_ap_b+q_aq_b) ,
\end{equation}
where 
\bea
H_{ab}&=&\int e_a[\phi]He_b[\phi]{\cal D}\phi,\eea
and the wave function is
\bea
\psi[\phi]&=&q[\phi]+ip[\phi],
\eea
where
\be q[\phi]=\sum_{a=1}^\infty q_a(t) e_a[\phi],
\qquad
p[\phi]=\sum_{a=1}^\infty p_a(t) e_a[\phi].
\ee

Thus, the classical Hamiltonian $ H_{sym}$ in quantum field theory has the same form as in non-relativistic quantum mechanics.

\section{Conclusion}
In this paper, the equivalence between the complex Hilbert space and real Kähler space formulations of quantum mechanics is  demonstrated.
This Kähler-space framework preserves all essential features of quantum mechanics. In particular,  the postulate for composite quantum systems can be formulated in the standard way,  and remains consistent with the corresponding postulate of real quantum mechanics in terms of Kähler spaces.
\\

 It is also demonstrated that the quantum mechanical Schrödinger equation is equivalent to a classical mechanical system with the Hamiltonian $H_{sym}$. The simple form of the Hamiltonian $H_{sym}$ is due to the fact that the Schrödinger equation, on the one hand, is a linear equation for a complex wave function, and, on the other hand, the corresponding linear equations can be rewritten for the real and imaginary parts of the wave function. Additionally, ergodicity in quantum theory can be treated as a classical phenomenon.
\\

Our results suggest that the Kähler formulation not only strengthens the conceptual connection between quantum and classical mechanics but also offers a computationally advantageous framework for exploring foundational questions in quantum theory.

\section*{Acknowledgment}
 I am grateful to  I.~Aref'eva, V. Kozlov, V.~Sakbaev, R. Singh, D.~Stepanenko, A.~Teretenkov and A.~Trushechkin for useful discussions.
This work is supported by the Russian Science Foundation (24-11-00039, Steklov Mathematical Institute).

\end{document}